\documentclass[
    ,final            
  ]
  {aipproc}

\layoutstyle{6x9}

\def\xv{{\bf x}}

\def\En{{\cal E}}
\def\f0{f_0}
\def\Lz{J_z}
\def\Q0{Q_0}
\def\Ra{R_{\rm a}}
\def\rhotil{\tilde\rho}
\def\sigR{\sigma_R}
\def\sigz{\sigma_z}
\def\sigphi{\sigma_{\varphi}}
\def\sigm{\sigma_{\rm m}}
\def\sigmtil{\tilde\sigm}
\def\sigphitil{\tilde\sigphi}
\def\vv{{\bf v}}

\def\vR{v_R}
\def\vz{v_z}
\def\vphi{v_{\varphi}}
\def\vm{v_{\rm m}}

\def\Rtil{{\tilde R}}

\def\Psit{\Psi_{\rm T}}
\def\Psiext{\Psi_{\rm ext}}
\def\phit{\phi_{\rm T}}

\begin{document}

\title{Self-consistent stellar dynamical tori}

\author{L. Ciotti}{
  address={Astronomy Department, Bologna University, Italy}}
\author{G. Bertin}{
  address={Physics Department, Milano University, Italy}}
\author{P. Londrillo}{
  address={INAF-Bologna Astronomical Observatory, Italy}}

\begin{abstract}
We present preliminary results on a new family of distribution
functions that are able to generate axisymmetric, truncated (i.e.,
finite size) stellar dynamical models characterized by {\it toroidal}
shapes. The relevant distribution functions generalize those that are
known to describe polytropic spheres, for which all the dynamical and
structural properties of the system can be expressed in explicit form
as elementary functions of the system gravitational potential. The
model construction is then completed by a numerical study of the
associated Poisson equation. We note that our axisymmetric models can
also include the presence of an external gravitational field, such as
that produced by a massive disk or by a central mass concentration
(e.g., a supermassive black hole).
\end{abstract}

\maketitle


\section{Introduction}

Constructing self-consistent collisionless equilibrium models is a key
step to understand the structure and the dynamics of stellar
systems. The construction process usually starts with the assignment
of a phase-space distribution function (DF) obeying the Jeans theorem
(and thus a solution of the collisionless Boltzmann equation that
describes such ``gravitational plasma''), which leads to an expression
for the density as a function of the potential; models are then
calculated by solving the Poisson equation, as a non-linear,
second-order partial differential equation for the potential. In the
presence of special symmetries (for example, in the case of spherical
models) the problem can be reduced to the study of an ordinary
differential equation, and the model construction is relatively
straightforward. For more general symmetries the procedure is
inherently difficult. In general, solutions have to be found
numerically and turn out to exist only for a finite range of the
parameters appearing in the adopted DF.

Here we present the basic properties of the simplest member of a new
family of DFs that are able to generate axisymmetric, truncated (i.e.,
finite size) stellar dynamical models characterized by {\it toroidal}
shapes.  The complete description of our family of DFs, together with
the results of N-body simulations aimed at the study of the {\it
stability} of these collisionless equilibrium configurations, will be
presented in a separate paper (Ciotti, Bertin, Londrillo 2004).  We
recall that toroidal, self-consistent models are sometimes mentioned
in the literature (e.g., see Lynden-Bell 1962; the Author expressed
concerns about their stability). With our models we will investigate
this matter further, also considering the possible stabilizing effect
of a massive dark matter halo. Astrophysical applications will address
the modeling of {\it peanut-shaped} bulges (e.g., see Aronica et
al. 2003, and references therein) and the problem of the {\it central
depression} recently discovered (Lauer et al. 2002) in the luminosity
profiles of some elliptical galaxies hosting supermassive black holes.

\section{The models}

The particular set of axisymmetric models presented here is
defined by the DF
\begin{equation}
f(Q,\Lz^2)=\f0 |\Lz|^{2\alpha}(Q-\Q0)^{\beta}\Theta(Q-\Q0),
\end{equation}
where $\f0$ is a (dimensional) physical scale, $Q\equiv\En
-\Lz^2/(2\Ra^2)$, $\Theta$ is the Heaviside step
function, and $\alpha$ and $\beta$ are two dimensionless constants;
the natural coordinates are cylindrical,
$(R,z,\varphi)$. In addition, $\En\equiv\Psit-||\vv||^2/2=
\Psit-(\vR^2+\vz^2+\vphi^2)/2$ is the binding energy per
unit mass, $\Lz=R\vphi$ is the axial component of the angular momentum
(per unit mass), and $\Q0$ and $\Ra$ are a {\it truncation energy} and
an {\it anisotropy radius}, respectively.  The function
$\Psit=\Psit(R,z)$ represents the total gravitational potential; for
the moment, we only require that $\Psit\sim {\rm O}(r^{-1})$ for
$r\to\infty$, where $r=\sqrt{R^2+z^2}$.  We allow for an external
component also, i.e., $\Psit=\Psi+\Psiext$, where $\Psi$ is the
potential associated with the density distribution derived from
eq. (1) as given by the following eq. (4), while $\Psiext$ is taken to
be a given function (equal to zero in the fully self-consistent case).
Note that we can write
\begin{equation}
Q=\Psit -{\vm^2\over 2}-{\vphi^2\over 2}\left(1+{R^2\over\Ra^2}\right),
\end{equation}
where $\vm^2\equiv\vR^2 +\vz^2$ is the square of the meridional (or
poloidal) velocity component.  Some dynamical properties of the
systems associated with eq. (1) are the following: $(i)$ no net
rotation is present, and $(ii)$ the two components $\sigR^2$ and
$\sigz^2$ of the velocity dispersion tensor are equal. In addition,
note that for $\alpha =0$ if we formally consider $\Ra\to\infty$ we
recover the polytropic spheres, while for integer $\alpha$ and $\beta$
the distribution function is a {\it sum} of Fricke's (1952) DFs. The
functional dependence of the system structural and dynamical
properties on $\Psit$ is obtained by integration over $\Omega(\xv)$,
the velocity section of phase--space over which $f\geq 0$ at given
$\xv$.  The coordinates $(v,\zeta,\xi)$ to be used to integrate the
various quantities over the rotationally symmetric (ellipsoidal)
region $\Omega(\xv)$ are naturally defined by
\begin{equation}
\vR=v\sin\zeta\cos\xi,\quad
\vz=v\sin\zeta\sin\xi,\quad
\vphi={v\cos\zeta\over\sqrt{1+R^2/\Ra^2}},
\end{equation}
where $0\leq\zeta\leq\pi$, $0\leq\xi\leq 2\pi$. With this choice, the
ellipsoidal radius $v$ of the velocity section $\Omega (\xv)$ runs in
the range $0\leq v\leq\sqrt{2[\Psit(\xv) -\Q0]}$. Integration over the
two angular coordinates gives
\begin{equation}
\rho = 4\pi\f0 g(R) 
         {B(\alpha+3/2,\beta+1)2^{\alpha+1/2}\over 2\alpha+1}
         (\Psit -\Q0)^{\alpha+\beta+3/2}\Theta(\Psit -\Q0),
\end{equation}
where $B(x,y)=\int_0^1 t^{x-1} (1-t)^{y-1}dt$ is the Euler Beta
function, 
\begin{equation}
g(R)= {R^{2\alpha}\over (1+R^2/\Ra^2)^{\alpha+1/2}},
\end{equation}
and for convergence it is required that $\alpha > -1/2$, $\beta
>-1$. Moreover,
\begin{equation} 
\rho\sigm^2 =4\pi\f0 g(R) 
               {B(\alpha+5/2,\beta+1)2^{\alpha +5/2}\over 
               (2\alpha +1)(2\alpha +3)}
               (\Psit -\Q0)^{\alpha+\beta+5/2}\Theta(\Psit -\Q0)
\end{equation}
and
\begin{equation}
\rho\sigphi^2 ={4\pi\f0 g(R)\over 1+R^2/\Ra^2}
                 {B(\alpha+5/2,\beta+1)2^{\alpha +3/2}\over 
                 2\alpha +3}
                 (\Psit -\Q0)^{\alpha+\beta+5/2}\Theta(\Psit -\Q0).
\end{equation}
The number of free parameters (excluding those associated with the
external potential) needed to describe a model completely is five: the
three dimensional quantities $\f0$, $\Ra$, and $\Q0$, and the two
dimensionless constants $\alpha$ and $\beta$. The relation between the
meridional velocity dispersion $\sigm^2$ and the azimuthal (toroidal)
velocity dispersion $\sigphi^2$, can be described by the anisotropy
distribution
\begin{equation}
a(R,z)\equiv 1-{2\sigphi^2\over \sigm^2}=
             {R^2/\Ra^2 -2\alpha\over R^2/\Ra^2 +1}.
\end{equation}
The velocity dispersion tensor is isotropic when $a=0$, tangentially
anisotropic when $a<0$, and meridionally anisotropic when $a>0$. Note
that the velocity dispersion anisotropy is constant on {\it
cylinders}. It can be proved that eq. (8) holds also when the factor
$(Q-\Q0)^{\beta}$ in eq. (1) is replaced by a generic function
$h(Q-\Q0)$ (Ciotti et al. 2004). We now impose the system consistency,
i.e., we require that
\begin{equation}
\triangle\Psi = -4\pi G\rho,
\end{equation}
with $\Psi$ a nowhere negative function.  In order to obtain the
numerical solution of eq. (9), the Poisson equation is first recast in
dimensionless form, by referring to the physical scales $\Ra$ for
lengths and $\Q0$ for potentials. After rescaling, we are then led to
the solution of the following problem:
\begin{equation}
\tilde\triangle\phi =-\lambda\rhotil,\quad\quad
                        \rhotil\equiv g(\Rtil)(\phit-1)^{\gamma}
                        \Theta (\phit-1),
\end{equation}
where $\gamma\equiv\alpha+\beta+3/2>0$, and the dimensionless
quantities are defined as $\phit\equiv\Psit/\Q0$, $\Rtil\equiv R/\Ra$,
and $g(\Rtil )\equiv \Rtil^{2\alpha}/(1+\Rtil^2)^{\alpha+1/2}$, with
\begin{equation}
\lambda\equiv {16\pi^2 G\f0\Ra^{2\alpha+2}\Q0^{\alpha+\beta+1/2} 
                B(\alpha+3/2,\beta+1)2^{\alpha +1/2}\over 2\alpha +1}.
\end{equation}
It follows that  
\begin{equation}
\rho={\Q0\over G\Ra^2}{\lambda\rhotil\over 4\pi}
\end{equation} 
and 
\begin{equation}
\sigmtil^2 ={2\over\gamma+1}(\phit-1)\Theta (\phit-1),\quad
\sigphitil^2 ={2\alpha+1\over (\gamma+1)}
                {(\phit-1)\Theta (\phit-1)\over 1+\Rtil^2}.
\end{equation}
\begin{figure}
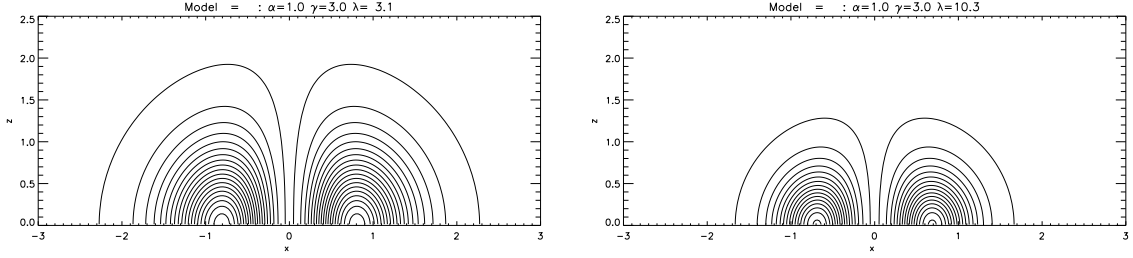

  \includegraphics[height=.25\textheight]{fig1.eps}
  \includegraphics[height=.25\textheight]{fig2.eps}
  \caption{Isodensity contours of a meridional section of the dimensionless
  density distribution associated with the models $(\alpha,\gamma)=(1,3)$ for
  $\lambda\simeq 3.1$ (left) and $\lambda\simeq 10.3$ (right).}
\end{figure}
Equation (10), a classical non-linear elliptic partial differential
equation, is then solved by using a Newton iteration scheme, described
in detail by Ciotti et al. (2004). In general, for assigned model
parameters we found convergence only for finite intervals of
$\lambda$. In Figure 1 we show the meridional sections of the density
distribution of two representative truncated and fully self-consistent
($\Psiext=0$) toroidal models, obtained by fixing $\alpha >0$ (see
eqs. [5] and [10]), for two different values of $\lambda$.

\begin{theacknowledgments}
The work of G.B. and L.C. has been partially supported by a grant
CoFin 2000 of the Italian MIUR.
\end{theacknowledgments}


\bibliographystyle{aipprocl} 

\end{document}